\documentclass[aps,prl,showpacs,twocolumn,amsmath,amssymb]{revtex4}
\usepackage{subfigure}
\usepackage{latexsym}
\usepackage{graphicx}
\usepackage{bm}

\newcommand{\ddt}[1]{\frac{\partial #1}{\partial t}}

\newcommand{\ddz}[1]{\frac{\partial #1}{\partial z}}

\newcommand{\bra}[1]{\langle #1 \vert}
\newcommand{\ket}[1]{\vert #1 \rangle}
\newcommand{\abs}[1]{\vert #1 \vert}

\begin{document}

\title{Stationary light pulses in ultra cold atomic gasses}

\author{Kristian Rymann Hansen}
\author{Klaus M\o lmer}
\affiliation{Lundbeck Foundation Theoretical Center for Quantum
System Research, Department of Physics and Astronomy, University of
Aarhus, DK-8000 \AA rhus C, Denmark}

\begin{abstract}
We present a theoretical treatment of electromagnetically induced
transparency and light storage using standing wave coupling fields
in a medium comprised of stationary atoms, such as an ultra cold
atomic gas or a solid state medium. We show that it is possible to
create stationary pulses of light which have a qualitatively
different behavior than in the case of a thermal gas medium,
offering greater potential for quantum information processing
applications.
\end{abstract}
 \maketitle


Electromagnetically induced transparency (EIT) \cite{Harris} in
ensembles of $\Lambda$ atoms has been extensively studied, both
experimentally
\cite{Liu,Phillips,Zibrov,Mair,Eisaman,Turukhin,Longdell} and
theoretically \cite{Lukin1,Mewes1,Mewes2,Peng} as a means for
coherent transfer of quantum states between photons and atoms.
Although the feasibility of using such systems for an efficient
quantum memory has been demonstrated, the nontrivial manipulation
of, and interaction between, stored quantum states remains a
challenge \cite{Lukin2} due to the weak interactions between the
atoms. Furthermore, the absence of any photonic component excludes
the use of enhanced nonlinear optical interactions
\cite{Schmidt,Harris2} between stored pulses. Several schemes for
using enhanced nonlinear optical interactions between slowly
propagating, weak light pulses have been proposed
\cite{Lukin3,Petrosyan,Ottaviani}, but the efficiency is limited
since the reduction in group velocity is accompanied by a reduction
in the energy of the pulse. Recent experimental progress in the
coherent control of light pulses in atomic media \cite{Bajcsy} has
demonstrated the possibility of generating stationary light pulses
by using EIT with a standing wave coupling laser in a thermal atomic
gas. It has been proposed \cite{Andre} that such a system can be
used for efficient nonlinear optical interactions between trapped
pulses. A theoretical treatment of stationary light pulses in a
thermal gas medium has been given in \cite{Zimmer}, but this theory
does not apply to media comprised of stationary atoms, such as an
ultra cold gas or a solid state crystal. In this Letter, we present
a theoretical treatment of stationary light pulses in the more
complicated case of a medium comprised of stationary atoms. It will
be shown that stationary light pulses trapped in such media display
a qualitatively different behavior and has greater potential for the
kind of nonlinear optical interactions envisaged in \cite{Andre}
than in the thermal gas case.

\begin{figure}
\includegraphics[width=\columnwidth]{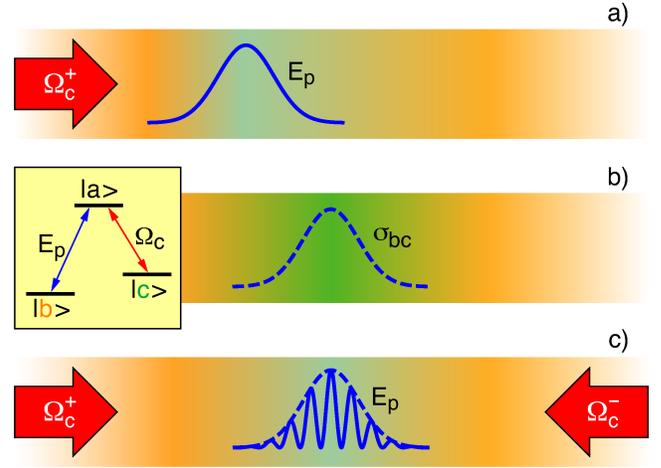}
\caption{\label{Fig:timing}(color online) Creation of a stationary
probe pulse: (a) The probe pulse propagates into the medium under
EIT conditions created by a traveling wave coupling laser. (b) The
probe pulse is stored in the medium by switching off the coupling
laser. (c) A stationary probe pulse is generated inside the medium
by switching on a standing wave coupling laser. Inset: Energy level
diagram for the three-level $\Lambda$ atom.}
\end{figure}

We consider a medium of length $L$ consisting of an ensemble of
stationary three-level atoms in the $\Lambda$ configuration (see
inset in Fig.~\ref{Fig:timing}) interacting with a weak quantum
field, called the probe field, and a strong classical field, called
the coupling field, propagating parallel to the $z$ axis. The probe
field is resonant with the transition between the ground state
$\ket{b}$ and the excited state $\ket{a}$, while the coupling field
is resonant with the transition between the metastable state
$\ket{c}$ and the excited state. The two lower states of the
$\Lambda$ atoms are assumed to be nearly degenerate, such that the
wave vectors of the probe and coupling fields are of equal
magnitude.

The probe field operator is written in terms of slowly varying
operators as
\begin{equation}
\hat{\mathbf{E}}_p(z,t)=\sqrt{\frac{\hbar\omega_p}{2\varepsilon_0V}}
\mathbf{e}_pE_p(z,t)e^{-i\omega_pt}+\text{h.c.}
\end{equation}
where $\omega_p$ is the carrier frequency of the probe field, $V$ is
the quantization volume and $\mathbf{e}_p$ is the polarization
vector. Since we are considering standing wave fields, the operator
$E_p$ is written as a superposition of two traveling wave fields,
denoted by $E_p^\pm$, propagating in opposite directions
\begin{equation}\label{Eq:probe_decomposition}
E_p(z,t)=E_p^+(z,t)e^{ikz}+E_p^-(z,t)e^{-ikz}.
\end{equation}

The $\Lambda$ atoms are described by collective atomic operators,
defined by
\begin{equation}
\hat{\sigma}_{\mu\nu}(z,t)=\frac{1}{N_z}\sum_{j=1}^{N_z}
\hat{\sigma}_{\mu\nu}^j(t),
\end{equation}
where $N_z$ is the number of atoms in a small volume centered around
$z$ and $\hat{\sigma}_{\mu\nu}^j=\ket{\mu_j}\bra{\nu_j}$. We
introduce temporally slowly varying operators for the coherences
\begin{subequations}
\begin{align}
\hat{\sigma}_{ba}(z,t)&=\sigma_{ba}(z,t)e^{-i\omega_pt}, \\
\hat{\sigma}_{ca}(z,t)&=\sigma_{ca}(z,t)e^{-i\omega_ct}, \\
\hat{\sigma}_{bc}(z,t)&=\sigma_{bc}(z,t)e^{-i(\omega_p-\omega_c)t},
\end{align}
\end{subequations}
where $\omega_p$ and $\omega_c$ are the carrier frequencies of the
probe and coupling lasers, respectively. The evolution of the
collective atomic operators is determined by Heisenberg-Langevin
equations. Assuming that the probe field is weak compared to the
coupling field, the Heisenberg-Langevin equations can be solved
perturbatively in the probe field. To first order, the relevant
Heisenberg-Langevin equations are
\begin{subequations}\label{Eq:Heisenberg-Langevin}
\begin{align}
\dot{\sigma}_{ba}&=i(g_pE_p+\Omega_c\sigma_{bc})
-\Gamma_{ba}\sigma_{ba} \\
\dot{\sigma}_{bc}&=i\Omega_c^*\sigma_{ba}-\Gamma_{bc}\sigma_{bc},
\label{Eq:Heisenberg-Langevin_b}
\end{align}
\end{subequations}
where $\Gamma_{ba}=\gamma_{ba}-i\delta_p$ and
$\Gamma_{bc}=\gamma_{bc}-i\Delta$ are the complex decay rates for
the optical coherence $\sigma_{ba}$ and the Raman coherence
$\sigma_{bc}$, which include the probe field detuning
$\delta_p=\omega_p-\omega_{ab}$ and the two-photon detuning
$\Delta=\omega_p-\omega_c-\omega_{cb}$. The vacuum Rabi frequency of
the probe field is defined by
$g_p=\sqrt{\frac{\omega_p}{2\hbar\varepsilon_0V}}
\mathbf{e}_p\cdot\mathbf{d}_{ba}$, where $\mathbf{d}_{ba}$ is the
dipole matrix element for the $a\leftrightarrow b$ transition, and
$\Omega_c$ is the Rabi frequency of the coupling field which has the
decomposition
\begin{equation}\label{Eq:coupling_decomposition}
\Omega_c(z,t)=\Omega_c^+(t)e^{ikz}+\Omega_c^-(t)e^{-ikz}.
\end{equation}
We have neglected the noise operators in the Heisenberg-Langevin
equations (\ref{Eq:Heisenberg-Langevin}) since it has been shown
\cite{Peng} that the effect of these is negligible in the adiabatic
limit we shall consider here.

The Heisenberg-Langevin equations (\ref{Eq:Heisenberg-Langevin}) can
be further simplified by assuming that the coupling laser Rabi
frequency and the probe field operator change slowly in time. By
introducing a characteristic timescale $T$ and expanding the
Heisenberg-Langevin equations in powers of $(\gamma_{ba}T)^{-1}$, we
find to lowest non-vanishing order
\begin{equation}\label{Eq:sigma_bc}
\sigma_{bc}=-\frac{g_pE_p}{\Omega_c}=-g_p
\frac{E_p^+e^{ikz}+E_p^-e^{-ikz}}{\Omega_c^+e^{ikz}+\Omega_c^-e^{-ikz}}.
\end{equation}
Inserting this expression into (\ref{Eq:Heisenberg-Langevin_b})
yields
\begin{equation}\label{Eq:sigma_ba}
\begin{split}
\sigma_{ba}&=\frac{-g_p\left(\Gamma_{bc}+\ddt{}\right)}{i\Omega
[1+2\abs{\kappa^+}\abs{\kappa^-}\cos(2kz+\phi)]} \\
&\quad\times\left(\frac{E_p^+e^{ikz}+E_p^-e^{-ikz}}{\Omega}\right),
\end{split}
\end{equation}
where we have introduced the time dependent total Rabi frequency
$\Omega(t)=\sqrt{\abs{\Omega_c^+}^2+\abs{\Omega_c^-}^2}$ and the
Rabi frequency ratios $\kappa^\pm=\frac{\Omega_c^\pm}{\Omega}$ which
are assumed to be constant. The phase angle $\phi$ is defined by the
relation $\kappa^+\kappa^{-*}=
\abs{\kappa^+}\abs{\kappa^-}e^{i\phi}$.

The evolution of the slowly varying probe field operators
(\ref{Eq:probe_decomposition}) in the slowly varying amplitude
approximation is governed by the wave equations
\begin{equation}\label{Eq:probe_waveeqn}
\left(\ddt{}\pm c\ddz{}\right)E_p^\pm(z,t)=ig_pN_z\sigma_{ba}^\pm
(z,t),
\end{equation}
where the two components $\sigma_{ba}^\pm$ of the optical coherence
with spatial dependence $e^{\pm ikz}$ are found by Fourier expanding
equation (\ref{Eq:sigma_ba}).

We now introduce new field operators, in analogy with the dark-state
polariton field introduced in \cite{Lukin1}, defined by
\begin{equation}\label{Eq:def_polariton}
E_p^\pm(z,t)=\cos\theta(t)\Psi^\pm(z,t),
\end{equation}
where the angle $\theta$ is given by $\tan\theta(t)=\frac{g_p
\sqrt{N_z}}{\Omega(t)}$.

Assuming that $\abs{\kappa^+}\geq\abs{\kappa^-}$ and considering the
low group velocity limit ($\cos^2\theta\ll 1$), we obtain a set of
coupled wave equations for the polariton field components
\begin{subequations}\label{Eq:polariton_waveeqn}
\begin{align}
\left(\Gamma_{bc}+\ddt{}\right)\Psi^++\abs{\kappa^+}^2v_g\ddz{\Psi^+}
&=\kappa^+\kappa^{-*}v_g\ddz{\Psi^-}, \\
\left(\Gamma_{bc}+\ddt{}\right)\Psi^--\abs{\kappa^+}^2v_g\ddz{\Psi^-}
&=-\kappa^{+*}\kappa^-v_g\ddz{\Psi^-}.
\end{align}
\end{subequations}

We consider the type of experiment performed by Bajcsy \emph{et al.}
\cite{Bajcsy} in which a probe pulse is stored as a Raman coherence
in the medium, using copropagating traveling wave lasers, and
subsequently retrieved by a standing wave coupling field (see timing
diagram in Fig.~\ref{Fig:timing}). Assuming that the standing wave
coupling field is switched on at $t=0$, the initial condition for
the Raman coherence of the atoms is
\begin{equation}
\sqrt{N_z}\sigma_{bc}(z,0)=-\Psi(z,0),
\end{equation}
where $\Psi(z,0)$ is a known function of $z$ found by solving the
traveling wave light storage problem covered in \cite{Lukin1}.
Inserting the initial condition for the Raman coherence into
(\ref{Eq:sigma_bc}) and using the definition of the polariton field
(\ref{Eq:def_polariton}), we obtain the initial conditions for the
two components $\Psi^\pm(z,0)$ of the polariton field
\begin{equation}
\Psi^+(z,0)=\kappa^+\Psi(z,0), \quad \Psi^-(z,0)=\kappa^-\Psi(z,0).
\end{equation}
With these initial conditions the solution to the wave equations
(\ref{Eq:polariton_waveeqn}) is
\begin{subequations}\label{Eq:polariton_solution}
\begin{align}
\begin{split}
\Psi^+(z,t)&=\frac{\kappa^+}{2}\biggl[\biggl(1+
\frac{\beta}{\abs{\kappa^+}^2}\biggr)\Psi(z-\beta r(t),0) \\
&\quad+\biggl(1-\frac{\beta}{\abs{\kappa^+}^2}\biggr)\Psi(z+\beta
r(t),0)\biggr]e^{-\Gamma_{bc}t}
\end{split} \\
\begin{split}
\Psi^-(z,t)&=\frac{\kappa^-}{2}\biggl(\Psi(z-\beta r(t),0) \\
&\quad+\Psi(z+\beta r(t),0)\biggr)e^{-\Gamma_{bc}t},
\end{split}
\end{align}
\end{subequations}
where
$\beta=\sqrt{\abs{\kappa^+}^2(\abs{\kappa^+}^2-\abs{\kappa^-}^2)}$
and
\begin{equation}
r(t)=\int_0^t c\cos^2\theta(t')\mathrm{d}t'.
\end{equation}
In the case of a standing wave coupling field
$(\kappa^+=\kappa^-=\frac{1}{\sqrt{2}})$, the solution becomes
\begin{subequations}
\begin{align}
\Psi^+(z,t)&=\frac{1}{\sqrt{2}}\Psi(z,0)e^{-\Gamma_{bc}t}, \\
\Psi^-(z,t)&=\frac{1}{\sqrt{2}}\Psi(z,0)e^{-\Gamma_{bc}t}.
\end{align}
\end{subequations}
Having found the solution for the polariton field, the solution for
the components of the probe field is easily obtained from equation
(\ref{Eq:def_polariton}).

The Raman coherence of the atoms is found from the adiabatic
solution (\ref{Eq:sigma_bc}). By inserting the decompositions
(\ref{Eq:probe_decomposition}) and (\ref{Eq:coupling_decomposition})
of the probe and coupling fields, as well as the definition
(\ref{Eq:def_polariton}) of the polariton field, we get
\begin{equation}\label{Eq:sigma_bc_expand}
\sqrt{N_z}\sigma_{bc}=-\sin\theta(t)
\frac{\Psi^+(z,t)e^{ikz}+\Psi^-(z,t)e^{-ikz}}{\kappa^+e^{ikz}+
\kappa^-e^{-ikz}}.
\end{equation}
It is clear that in the case of a standing wave coupling field
$(\abs{\kappa^+}=\abs{\kappa^-})$, only the dc Fourier component is
present,
\begin{equation}
\sqrt{N_z}\sigma_{bc}^{(0)}=-\sin\theta\Psi(z,0)e^{-\Gamma_{bc}t},
\end{equation}
while in the more general case of a quasi-standing wave coupling
field ($\abs{\kappa^+}>\abs{\kappa^-}$), we find by a binomial
expansion of equation (\ref{Eq:sigma_bc_expand})
\begin{equation}
\begin{split}
\sqrt{N_z}\sigma_{bc}&=-\frac{1}{2}\sin\theta\biggl(\Psi(z-\beta
r,0)+\Psi(z+\beta r,0) \\
&\quad+\frac{\beta}{\abs{\kappa^+}^2}[\Psi(z-\beta r,0)-\Psi(z+\beta
r,0)] \\
&\quad\times\sum_{n=0}^\infty\left(-\frac{\kappa^-}{\kappa^+}
\right)^n e^{-2inkz}\biggr)e^{-\Gamma_{bc}t}.
\end{split}
\end{equation}
We observe that $\sigma_{bc}$ has Fourier components $e^{-2inkz}$
with only positive $n$, becoming progressively smaller with
increasing $n$.

\begin{figure}
\subfigure[\ Cold gas]
{\includegraphics[width=4cm]{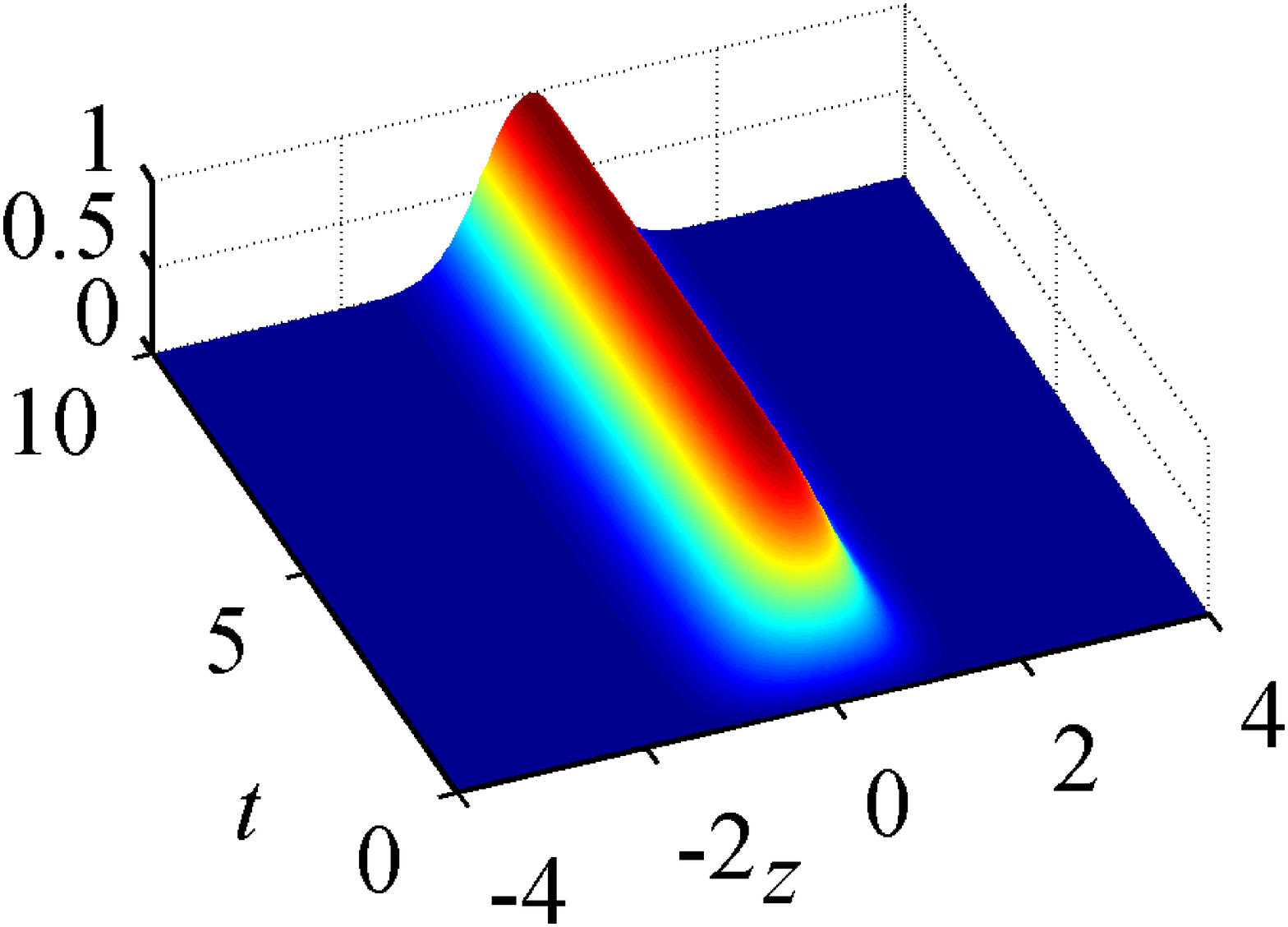}}\label{Fig:standa}
\subfigure[\ Thermal gas]
{\includegraphics[width=4cm]{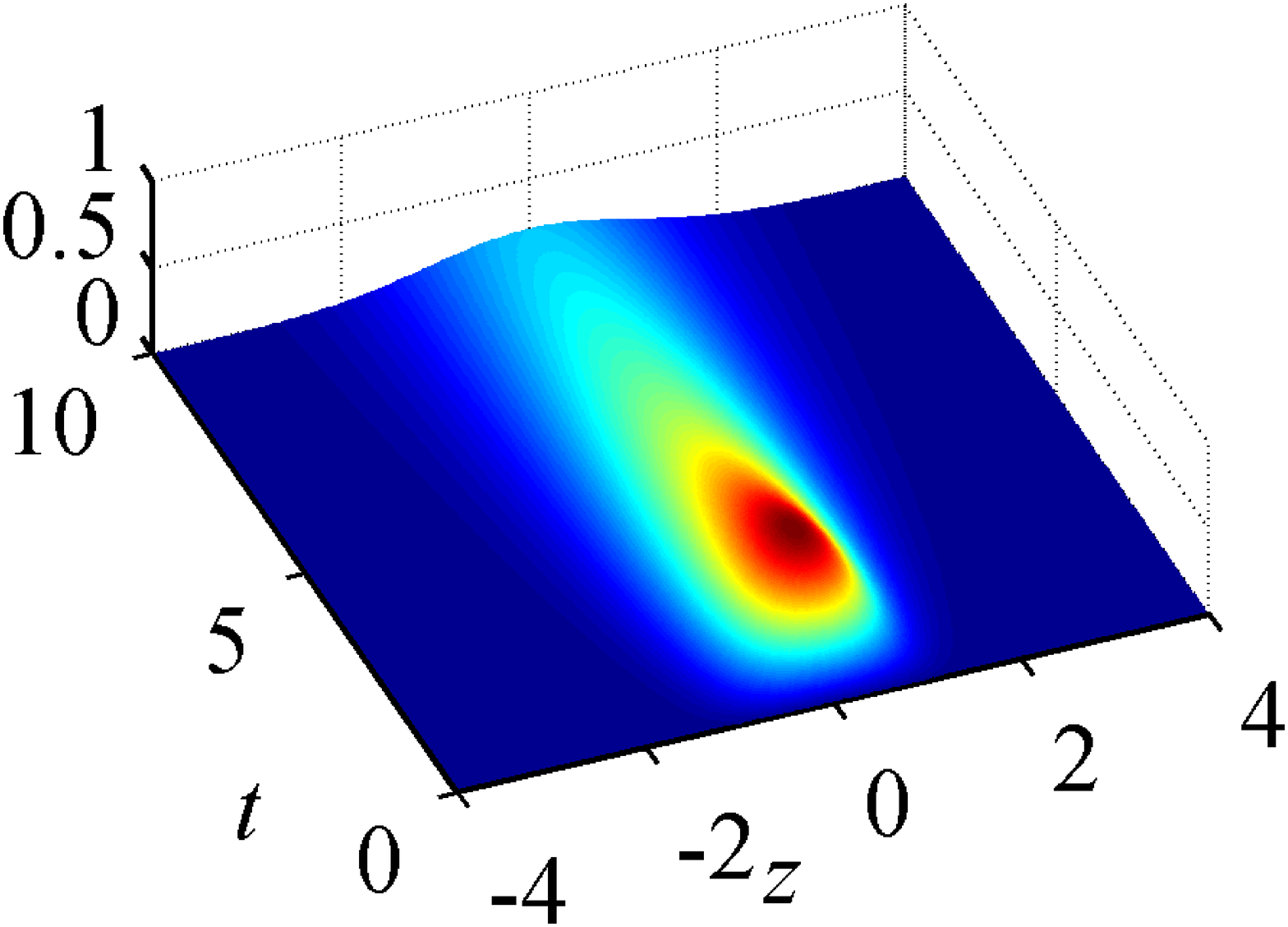}}\label{Fig:standb}
\caption{\label{Fig:stand}(color online) Retrieval of a stored probe
pulse with a standing wave coupling field. The normalized probe
field energy density is plotted for a medium comprised of: (a) cold
atoms, (b) thermal atoms. Time $t$ is in units of the switching time
$T_s$, position $z$ is in units of the pulse length
$L_p=v_{g,0}T_s$. The absorption length of the media is taken to be
$l_a=0.1\times L_p$.}
\end{figure}

Comparing our solution (\ref{Eq:polariton_solution}) for the
stationary atom case to the solution for the thermal gas case
presented in \cite{Zimmer}, we find significant qualitative and
quantitative differences. First, the diffusive broadening of the
retrieved probe pulse seen in the thermal gas case is absent in a
medium comprised of stationary atoms. This makes such media ideally
suited for quantum information processing applications, since the
dissipative losses associated with the broadening are also absent.
In Fig.~\ref{Fig:stand} the retrieval of a stored probe pulse by a
standing wave coupling field in ultra cold atomic gasses and in
thermal atomic gasses are compared. As an example, the initial
conditions for the components $\Psi^\pm$ of the polariton field are
given by the function $\Psi(z,0)=\Psi_0\exp(-(z/L_p)^2)$ and the
time dependence of the coupling field is given by
$\cos^2\theta(t)=\cos^2\theta_0\tanh(t/T_s)$ for $t\geq 0$, where
$T_s$ is the characteristic switching time. We have taken the length
of the stored probe pulse to be $L_p=v_{g,0}T_s$, where
$v_{g,0}=c\cos^2\theta_0$ is the group velocity of the polariton
field prior to storage, and assumed negligible ground state
dephasing ($\Gamma_{bc}=0$). The absorption length of both media in
the absence of EIT is taken to be $l_a=0.1\times L_p$. In actual
experiments, the dephasing rate $\gamma_{bc}$ ranges from a few kHz
in atomic gasses \cite{Liu,Phillips} to a few tens of kHz in solid
state media \cite{Longdell}, allowing storage times much longer than
the typical temporal length of the probe pulse, which is of the
order of a few $\mu$s. The inclusion of a non-zero dephasing rate
would therefore not significantly alter the results presented here.

\begin{figure}
\subfigure[\
$\Psi^+(z,t)$]{\includegraphics[width=4cm]{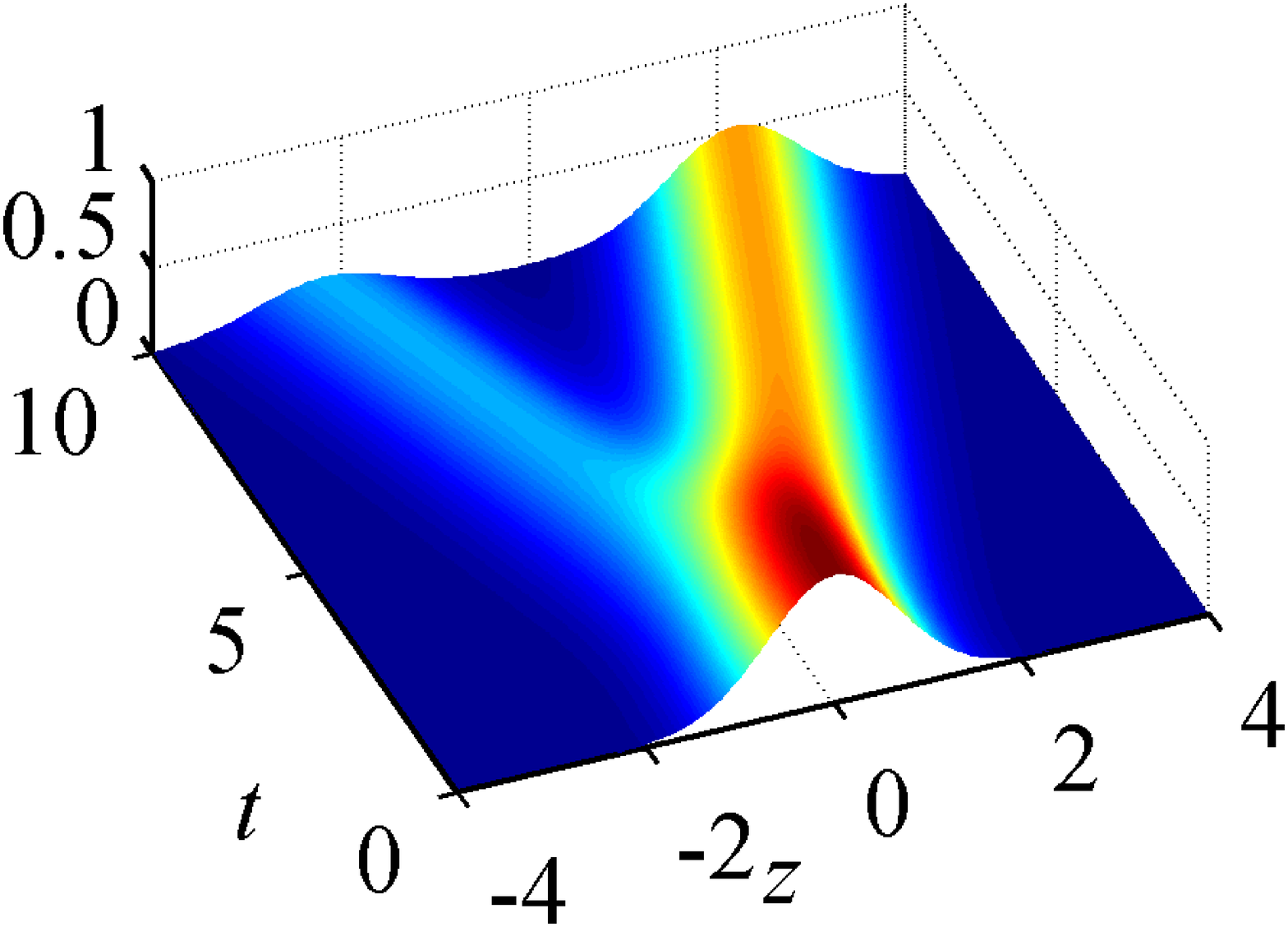}
\label{Fig:qstand1a}} \subfigure[\
$\Psi^+(z,t)$]{\includegraphics[width=4cm]{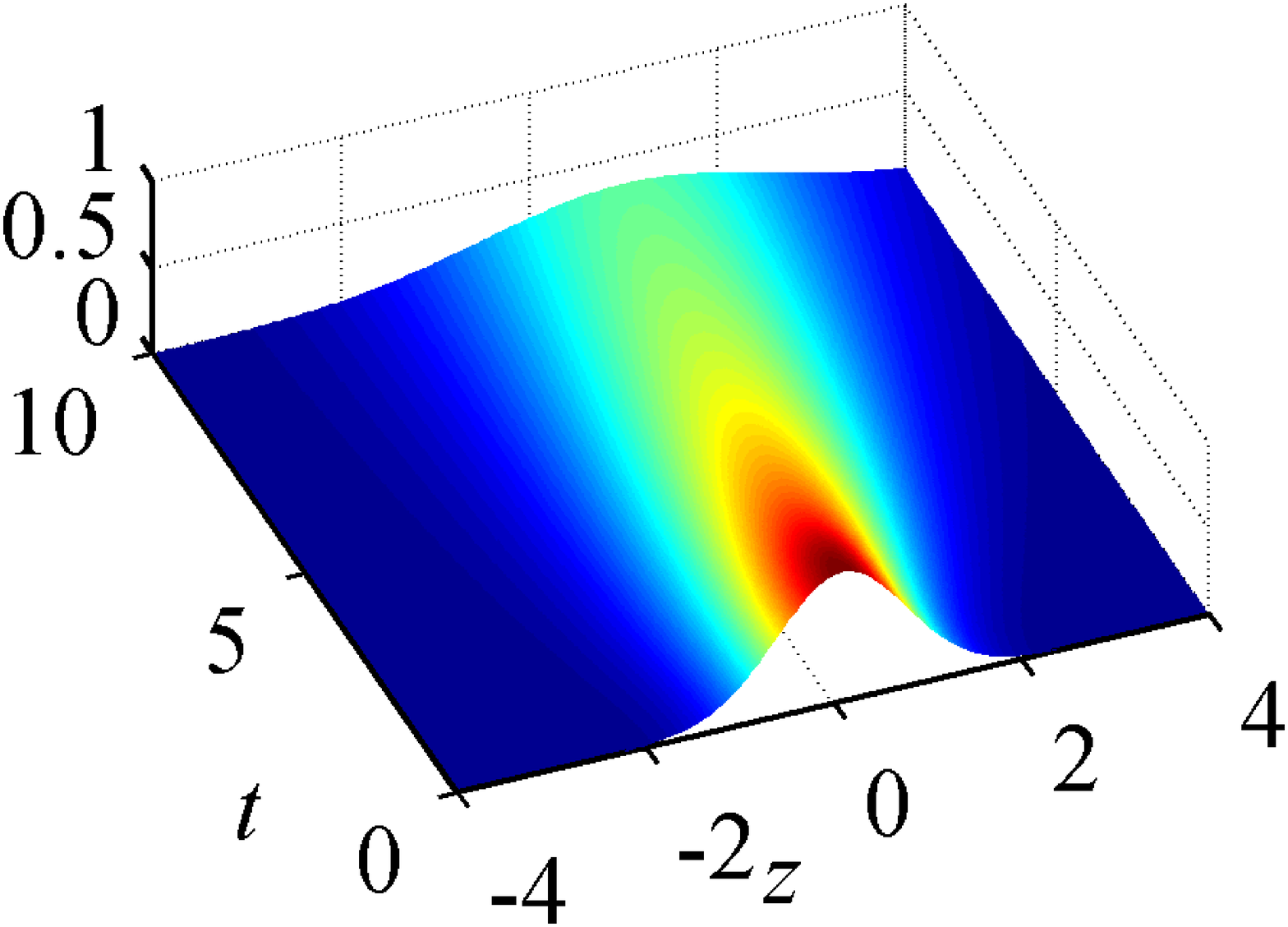}
\label{Fig:qstand1b}} \\
\subfigure[\
$\Psi^-(z,t)$]{\includegraphics[width=4cm]{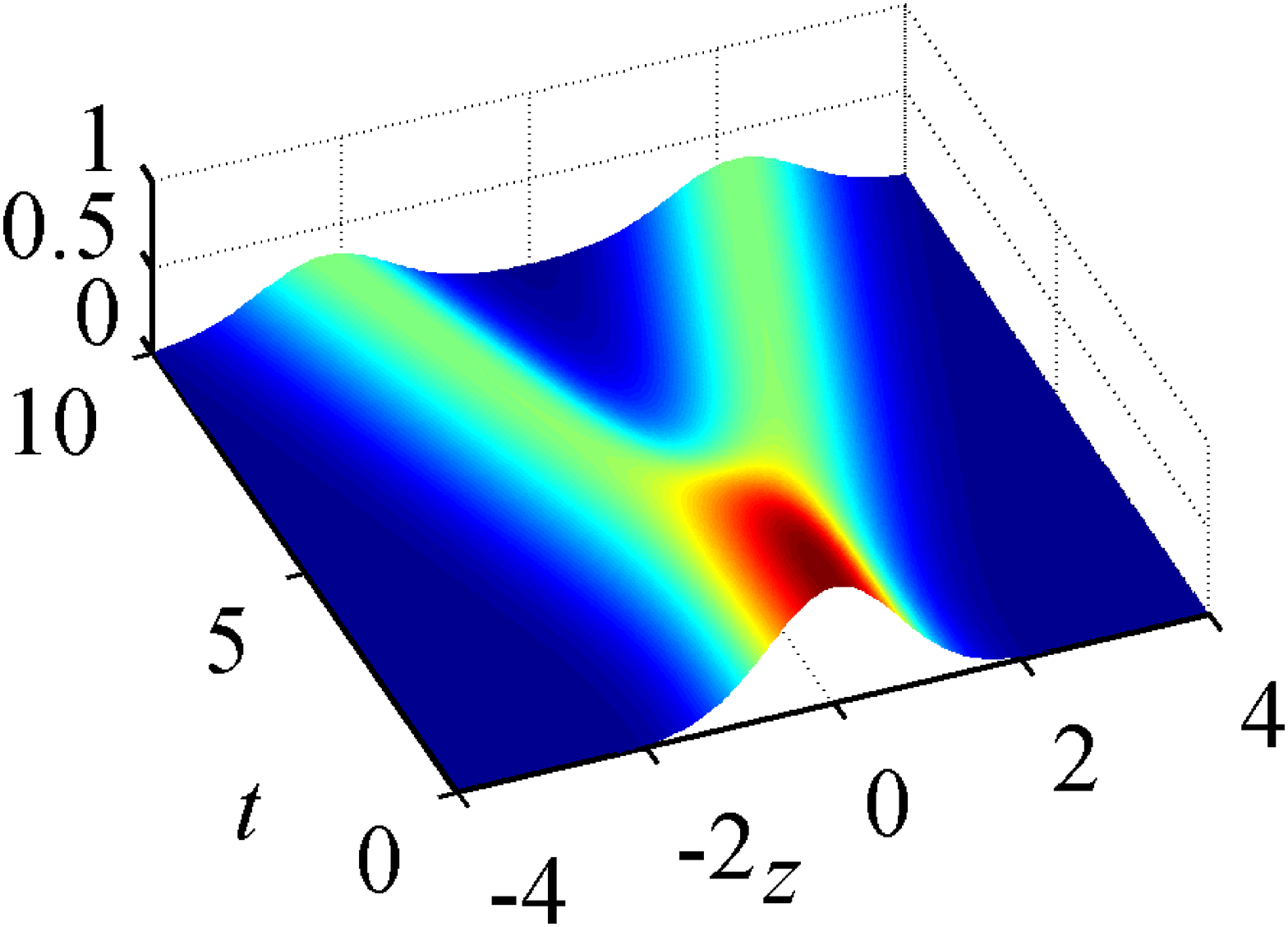}
\label{Fig:qstand1c}} \subfigure[\
$\Psi^-(z,t)$]{\includegraphics[width=4cm]{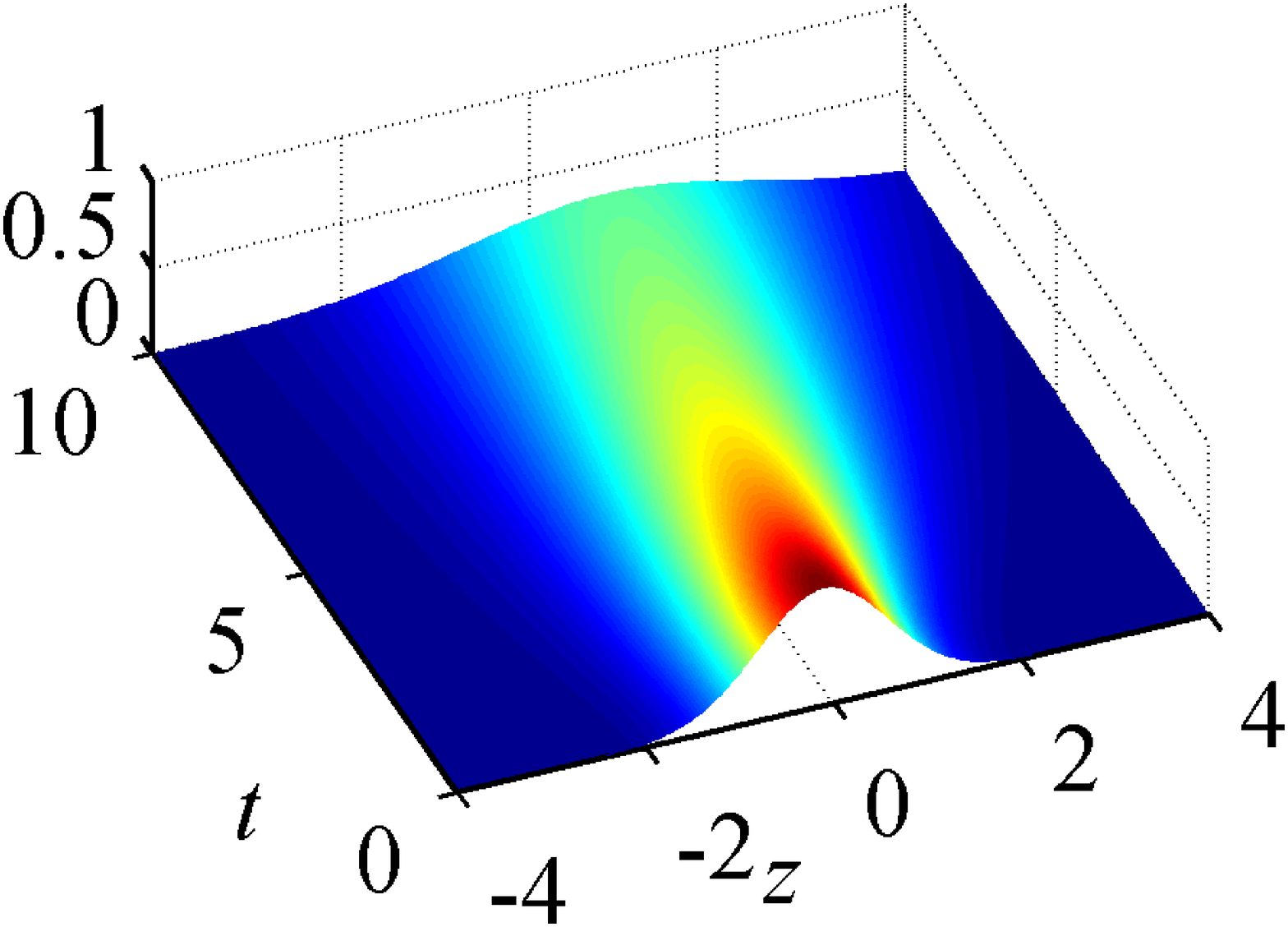}
\label{Fig:qstand1d}} \\
\subfigure[\
$\abs{E_p^+}^2+\abs{E_p^-}^2$]{\includegraphics[width=4cm]{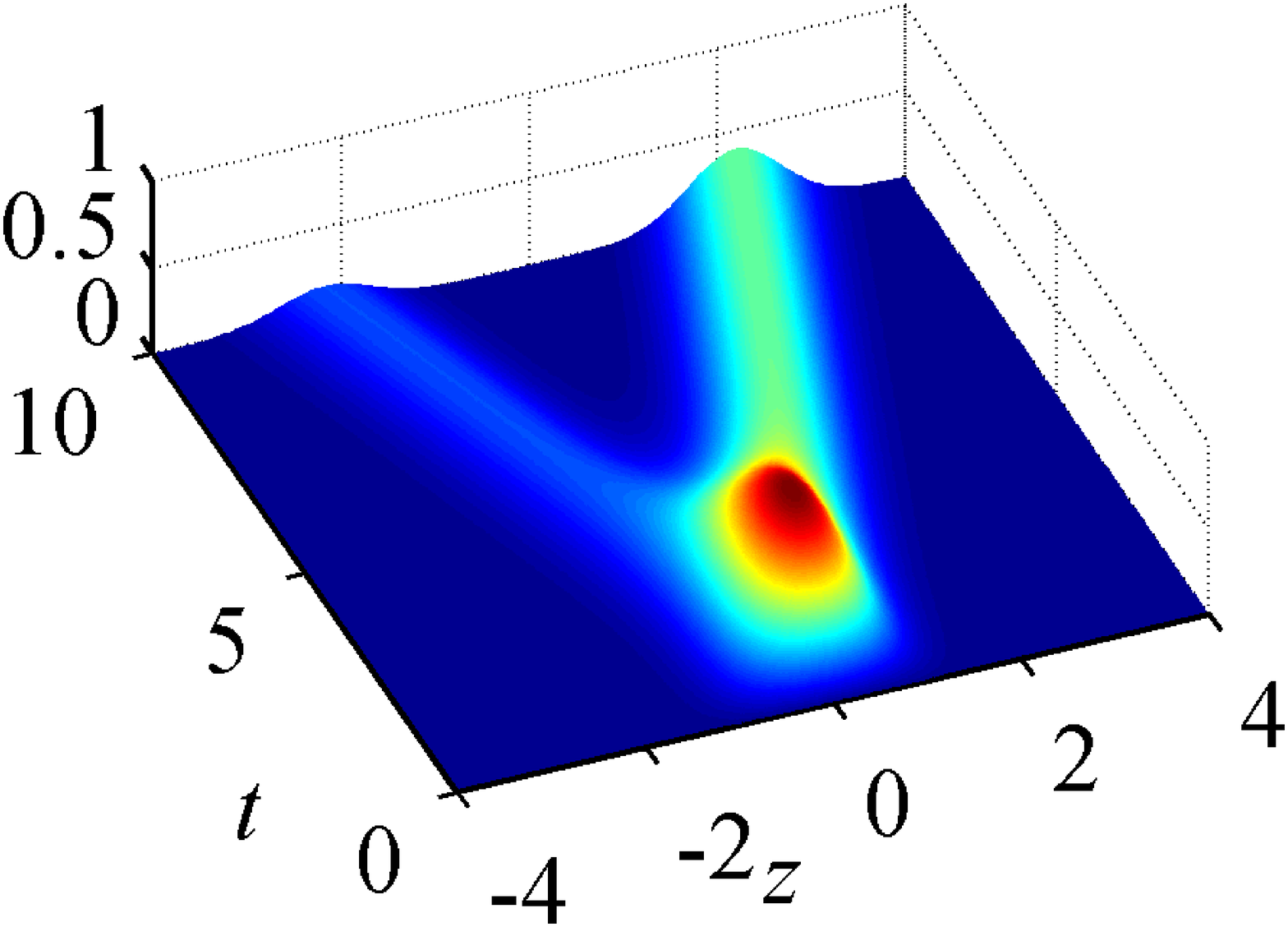}}
\label{Fig:qstand1e} \subfigure[\
$\abs{E_p^+}^2+\abs{E_p^-}^2$]{\includegraphics[width=4cm]{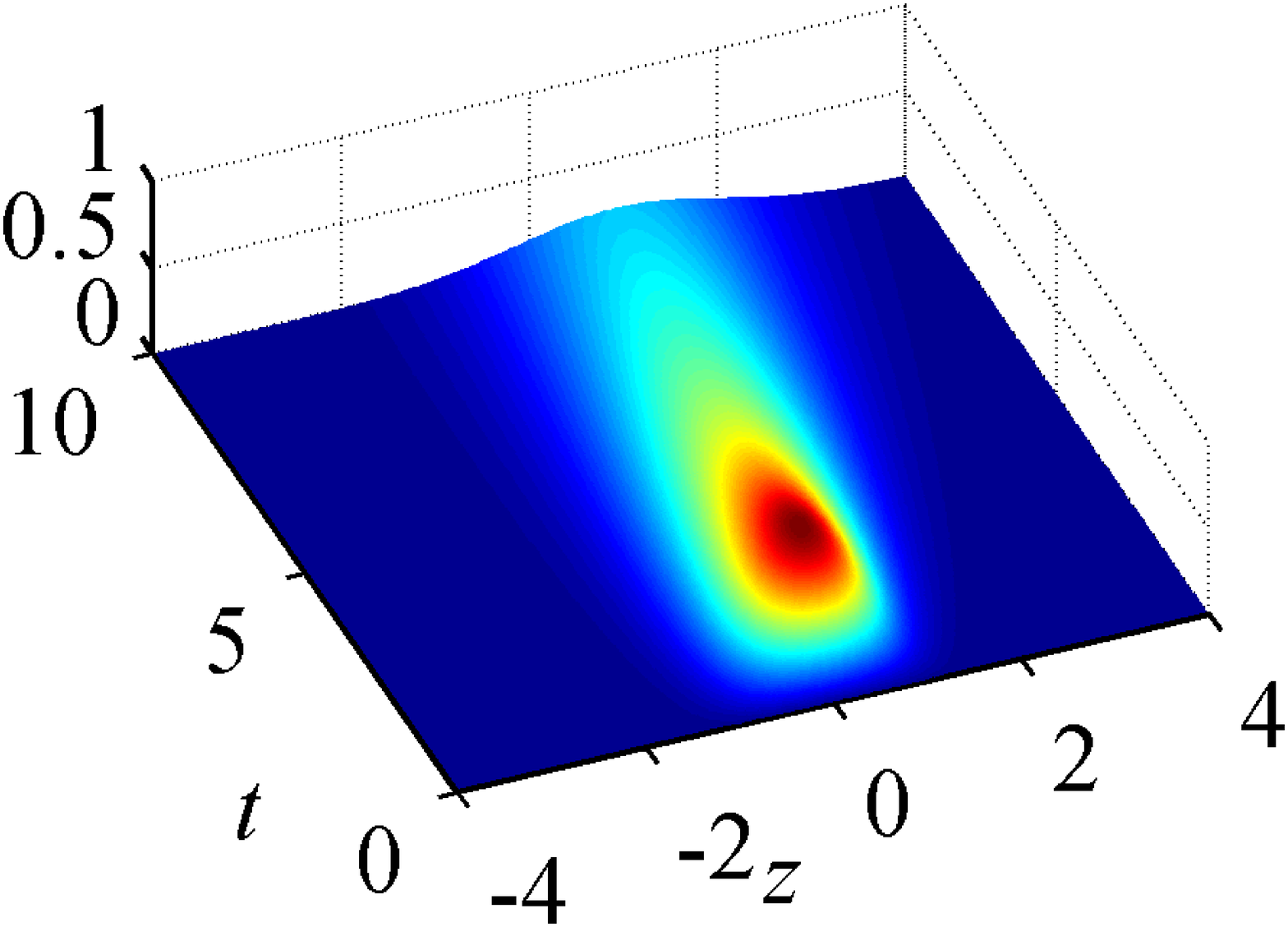}}
\label{Fig:qstand2b}\caption{\label{Fig:qstand1}(color online)
Retrieval of a stored probe pulse with a quasi-standing wave
coupling field ($\kappa^+=\sqrt{0.55}$, $\kappa^-=\sqrt{0.45}$). The
components of the polariton field and the probe field intensity is
shown for both the ultra cold gas case (left column) and for the
thermal gas case (right column).}
\end{figure}

Secondly, the behavior of a probe field retrieved by a
quasi-standing wave coupling field is very different in the
stationary atom case. In Fig.~\ref{Fig:qstand1} the retrieval of a
stored probe pulse by a quasi-standing wave coupling field with
$\kappa^+=\sqrt{0.55}$, $\kappa^-=\sqrt{0.45}$ in ultra cold atomic
gasses and thermal atomic gasses are compared. It was shown in
\cite{Zimmer} that a probe field retrieved by a quasi-standing wave
coupling field in a thermal gas medium acquires a small group
velocity $v_g=(\abs{\kappa^+}^2 -\abs{\kappa^-}^2)c\cos^2\theta$.
Our solution (\ref{Eq:polariton_solution}) shows that in the case of
stationary atoms, the revived probe field splits into two distinct
parts: a stronger part propagating in the direction of the stronger
component of the coupling field and a weaker part propagating in the
opposite direction with velocities $v_g=\pm\sqrt{\abs{\kappa^+}^2
(\abs{\kappa^+}^2-\abs{\kappa^-}^2)}c\cos^2\theta$.
We attribute the difference between moving and stationary atoms to
the presence of the spatially rapidly varying Fourier components of
the Raman coherence in the stationary atom case. These components
undergo rapid dephasing due to the motion of the atoms, and are
therefore suppressed in a thermal gas medium.

In summary, we have developed a theoretical treatment of light
storage and retrieval using standing wave coupling lasers in a
medium comprised of stationary atoms, such as an ultra cold gas or a
solid state medium. We find that the diffusive broadening of the
probe pulse seen in a thermal gas medium is absent in a medium
comprised of stationary atoms, and that the behavior of a probe
pulse retrieved by a quasi-standing wave coupling field is
significantly different in the two types of media. These differences
have important consequences for experiments and applications of
slowly propagating and stationary light. For example, the splitting,
rather than slow propagation, of the probe field in a quasi-standing
wave (see Fig.~\ref{Fig:qstand1}) prevents the nonlinear optical
interaction proposed in \cite{Andre} between a slowly propagating
probe pulse and a stored polariton in media comprised of stationary
atoms. Our results suggest to implement the opposite protocol with a
stationary probe pulse interacting with a slowly propagating
polariton in such media. The longer interaction times and stronger
interactions possible in the absence of diffusive broadening (see
Fig.~\ref{Fig:stand}) make media comprised of stationary atoms
ideally suited for quantum information processing applications.



\begin{thebibliography}{99}
\bibitem{Harris} S. E. Harris, Phys. Today $\mathbf{50}$(7), 36 (1997)
\bibitem{Liu} C. Liu, Z. Dutton, C. H. Behroozi, L. V. Hau, Nature
$\mathbf{409}$, 490 (2001)
\bibitem{Phillips} D. F. Phillips, A. Fleischhauer, A. Mair, R. L.
Walsworth, M. D. Lukin, Phys. Rev. Lett. $\mathbf{86}$, 783 (2001)
\bibitem{Zibrov} A. S. Zibrov, A. B. Matsko, O. Kocharovskaya, Y. V.
Rostovtsev, G. R. Welch, M. O. Scully, Phys. Rev. Lett.
$\mathbf{88}$, 103601 (2002)
\bibitem{Mair} A. Mair, J. Hager, D. F. Phillips, R. L. Walsworth,
M. D. Lukin, Phys. Rev. A $\mathbf{65}$, 031802 (2002)
\bibitem{Eisaman} M. D. Eisaman, L. Childress, A. Andr\'e, F. Massou,
A. S. Zibrov, M. D. Lukin, Phys. Rev. Lett. $\mathbf{93}$, 233602
(2004)
\bibitem{Turukhin} A. V. Turukhin, V. S. Sudarshanam, M. S.
Shahriar, J. A. Musser, B. S. Ham, P. R. Hemmer, Phys. Rev. Lett.
$\mathbf{88}$, 023602
\bibitem{Longdell} J. J. Longdell, E. Fraval, M. J. Sellars, N. B.
Manson, Phys. Rev. Lett. $\mathbf{95}$, 063601 (2005)
\bibitem{Lukin1} M. Fleischhauer, M. D. Lukin, Phys. Rev. Lett.
$\mathbf{84}$, 5094 (2000); M. Fleischhauer, M. D. Lukin, Phys. Rev.
A $\mathbf{65}$, 022314 (2002)
\bibitem{Mewes1} C. Mewes, M. Fleischhauer, Phys. Rev. A
$\mathbf{66}$, 033820 (2002)
\bibitem{Mewes2} C. Mewes, M. Fleischhauer, Phys. Rev. A
$\mathbf{72}$, 022327 (2005)
\bibitem{Peng} A. Peng, M. Johnsson, W. P. Bowen, P. K. Lam, H.-A. Bachor, J. J. Hope,
Phys. Rev. A $\mathbf{71}$, 033809 (2005)
\bibitem{Lukin2} M. D. Lukin, Rev. Mod. Phys. $\mathbf{75}$, 457 (2003)
\bibitem{Schmidt} H. Schmidt, A. Imamo\u glu, Opt. Lett.
$\mathbf{21}$, 1936 (1997)
\bibitem{Harris2} S. E. Harris, L. V. Hau, Phys. Rev. Lett.
$\mathbf{82}$, 4611 (1999)
\bibitem{Lukin3} M. D. Lukin, A. Imamo\u glu, Phys. Rev. Lett.
$\mathbf{84}$, 1419 (2000)
\bibitem{Petrosyan} D. Petrosyan, G. Kurizki, Phys. Rev. A
$\mathbf{65}$, 033833 (2002)
\bibitem{Ottaviani} C. Ottaviani, D. Vitali, M. Artoni, F.
Cataliotti, P. Tombesi, Phys. Rev. Lett. $\mathbf{90}$, 197902
(2003)
\bibitem{Bajcsy} M. Bajcsy, A. S. Zibrov, M. D. Lukin, Nature
$\mathbf{426}$, 638 (2003)
\bibitem{Andre} A. Andr\'e, M. Bajcsy, A. S. Zibrov, M. D. Lukin,
Phys. Rev. Lett. $\mathbf{94}$, 063902 (2005)
\bibitem{Zimmer} F. E. Zimmer, A. Andr\'e, M. D. Lukin, M.
Fleischhauer, Opt. Commun. $\mathbf{264}$, 441 (2006)
\end{thebibliography}
\end{document}